\documentclass[twocolumn,english,conference]{IEEEtran}
\ifCLASSINFOpdf
\else
\fi

\usepackage[T1]{fontenc}
\usepackage{babel}
\usepackage{array}
\usepackage{calc}
\usepackage{amsmath}
\usepackage{amssymb}
\usepackage{graphicx}
\usepackage{cite}
\usepackage{algorithm}
\usepackage{subfigure}
\usepackage{color}

\usepackage{enumitem}
\usepackage{multirow}
\usepackage{hhline}

\usepackage[mathlines,switch]{lineno}

\usepackage[unicode=true,
bookmarks=true,bookmarksnumbered=true,bookmarksopen=true,bookmarksopenlevel=1,
breaklinks=false,pdfborder={0 0 0},backref=false,colorlinks=false]
{hyperref}
\hypersetup{pdftitle={Your Title},
	pdfauthor={Your Name},
	pdfpagelayout=OneColumn, pdfnewwindow=true, pdfstartview=XYZ, plainpages=false}

\makeatletter

\providecommand{\tabularnewline}{\\}

\let\oldforeign@language\foreign@language
\DeclareRobustCommand{\foreign@language}[1]{%
	\lowercase{\oldforeign@language{#1}}}

\makeatother
\allowdisplaybreaks

\makeatletter
\newcommand{\thickhline}{%
	\noalign {\ifnum 0=`}\fi \hrule height 1.5pt
	\futurelet \reserved@a \@xhline
}
\newcolumntype{"}{@{\hskip\tabcolsep\vrule width 1pt\hskip\tabcolsep}}
\makeatother

\begin{document}
	%
	\title{Evaluating Power System Vulnerability to False Data Injection Attacks via Scalable Optimization}

	
	
	%
	\author{\IEEEauthorblockN{Zhigang Chu,
			Jiazi Zhang,
			Oliver Kosut, and
			Lalitha Sankar}
		\IEEEauthorblockA{School of Electrical, Computer, and Energy Engineering\\
			Arizona State University\\
			Tempe, AZ, 85287}}


	\maketitle
\pagestyle{plain}
	\begin{abstract}
		Physical consequences to power systems of false data injection cyber-attacks are considered. Prior work has shown that the worst-case consequences of such an attack can be determined using a bi-level optimization problem, wherein an attack is chosen to maximize the physical power flow on a target line subsequent to re-dispatch. This problem can be solved as a mixed-integer linear program, but it is difficult to scale to large systems due to numerical challenges. Three new computationally efficient algorithms to solve this problem are presented. These algorithms provide lower and upper bounds on the system vulnerability measured as the maximum power flow subsequent to an attack. Using these techniques, vulnerability assessments are conducted for IEEE 118-bus system and Polish system with 2383 buses.
	\end{abstract}
	

	%
	\IEEEpeerreviewmaketitle
	\global\long\def\figurename{Fig.}
	\global\long\def\tablename{TABLE}

	\section{Introduction}
	With integration of real-time monitoring, sensing, communication and data processing, electric power systems are becoming increasingly efficient and intelligent. However, similar to all computer network integrated systems, this integration also makes power systems more vulnerable to cyber attacks which could result in serious physical consequences or even system failure. Therefore, it is crucial to evaluate system vulnerability to credible attacks before they happen, and develop techniques to detect potential attacks and protect the system.
	
	False data injection (FDI) involves a malicious adversary replacing a subset of measurements with counterfeits. It has been shown that FDI attacks can be designed to target system states \cite{Liu2009}, \cite{Kosut2011}, \cite{Hug2012}, system topology \cite{Kim2013a}, \cite{Jzhang2016}, generator dynamics \cite{Kundur2010}, and energy markets \cite{Xie2011}. Some of these involve designing an optimization problem to determine the worst-case FDI attacks that can cause line overflow \cite{Liang2015}, operating cost change \cite{Yuan11}, \cite{Yuan2012}, or locational marginal price change \cite{Jia2014}. However, the results are only demonstrated for small systems. Similar to \cite{Liang2015}, we consider an optimization problem to determine the worst-case FDI attack that causes line overflow, but our goal is to design optimization algorithms that scale to significantly larger systems (\emph{i.e.} thousands of buses).
	
	
	In \cite{Liang2015}, an FDI attack against state estimation (SE) that leads to an overflow is introduced. Subsequent to the attack, the system operator re-dispatches the system generation, leading to an overload on a target line. Modeling such attacks leads to formulation of a bi-level optimization problem in which the first level models the attacker's ability and limitations while the second level models the system response via optimal power flow (OPF). This bi-level optimization problem is re-formulated to a single level mixed-integer linear problem (MILP) by replacing the second level by its Karush-Kuhn-Tucker (KKT) conditions and rewriting the non-convex complementary slackness conditions as mixed integer constraints. 
	
	As the system size scales, this optimization problem becomes hard to solve because of the increasing number of constraints as well as number of binary variables. But the recent cyber attack in Ukraine (see \cite{UkraineAttack}) reminds us that it is imperative to characterize the vulnerability of large power systems. To this end, we introduce three computationally efficient algorithms to characterize the vulnerability of systems. In some cases, these algorithms give the optimal attack. In other cases, finding the optimal attack is intractable, so instead these algorithms provide lower and upper bounds on the optimal objective value. A lower bound represents a feasible solution, and thus, highlights a specific overflow vulnerability for the system. An upper bound, on the other hand, constitutes a limit on the severity of this class of attacks. 
	 
	The first algorithm provides the optimal solution to the problem by reducing the number of line thermal limit constraints as well as the binary variables associated with them. The second algorithm further reduces the number of generation limit constraints and corresponding binary variables. However, by doing this we give up a guarantee on optimality, so the algorithm only provides a lower bound on the objective. The third algorithm provides both a lower bound and an upper bound via linear programming (LP) that maximizes the difference between target line cyber and physical power flows. All three algorithms are tested on the IEEE 118-bus system and the Polish system (2383 buses) to evaluate system vulnerability. 
	
	The outline of this paper is as follows. In Sec. \ref{sec: SE and models}, power system state estimation and the attack model are described. Sec. \ref{sec:OldAttack} describes how to formulate the bi-level optimization problem and how to convert it to a MILP, while Sec. \ref{sec:Methods} introduces three algorithms to solve such optimization problem. Simulation results and concluding remarks are presented in Sec. \ref{sec:Simulation} and \ref{sec:conclusion}, respectively.
	
	
	\section{\label{sec: SE and models}State Estimation and Attack Model}
	In this section, we introduce the mathematical formulation for SE and the attack model. Throughout, we assume there are $n_b$ buses, $n_{br}$ branches, $n_g$ generators, and $n_m$ measurements in the system. For tractability, we focus on the DC power flow model and DC SE, but the attacks introduced in this paper can also be performed against AC SE as in \cite{Liang2015}.
	
	\subsection{State Estimation}
	The DC measurement model is described as
	\begin{equation}
		z=Hx+e\label{eq:DCMeasurement}
	\end{equation}
	where $z$ is the $n_{m}\times1$ measurement vector whose entries are measurements of the system; $x$ is the vector of bus voltage angles; $H$ denotes the $n_{m}\times n_{b}$ matrix describing the relationship between the system states and measurements; $e$ is the $n_{m}\times1$ measurement error vector, whose entries are assumed to be independent and Gaussian distributed with zero mean and covariance matrix $R=\text{diag}(\sigma_{1}^{2},\sigma_{2}^{2},...,\sigma_{n_{m}}^{2})$.
	
	Observability analysis is performed before the state estimation process to check whether the system is fully observable. The weighted least-square (WLS) method is utilized to solve the SE problem, and the solution is given by \cite{Liu2009}
	\begin{equation}
		\hat{x}=(H^{T}R^{-1}H)^{-1}H^{T}R^{-1}z\label{eq: WLS}
	\end{equation}
	where $\hat{x}$ is the estimated system state vector. We assume classical bad data detection, based on measurement residuals, is used to detect large errors in measurement data. Note that traditional bad data detectors cannot necessarily detect FDI attacks. Indeed, unobservable attacks, as defined below, cannot be detected by any bad data detector based on measurement residuals.
	\subsection{\label{AttackModel}Attack Model}
	We first assume that the attacker has the following knowledge and capabilities:
	\begin{enumerate}
		\item The attacker has full system topology information via power transfer distribution factors (PTDF).
		\item The attacker has knowledge of load distribution, generation costs and line thermal limits of the system. 
		\item The attacker has control of the measurements in a subset $\mathcal{S}$ of the network.
	\end{enumerate}
	
	As discussed in \cite{Liang2015}, in the absence of noise, an attack is defined to be \emph{unobservable} when there exists an $n_{b}\times 1$ attack vector $c \neq 0$ such that for all $i$, the measurement $\bar{z}_{i}$ modified by the attacker satisfies $\bar{z}_{i}=z_{i}+H_{i}c$, where $H_{i}$ denotes the $i^{\text{th}}$ row of $H$. Given an attacker with control of the measurements in $\mathcal{S}$, it can execute this attack with attack vector c if $H_i c$ has non-zero entries only in $\mathcal{S}$.
	
	Given an attack vector $c$, the following procedure produces a subgraph $\mathcal{S}$ that, if controlled by the attacker, can execute an unobservable attack. For an attack vector $c$, \textit{load buses} (\emph{i.e.}, buses with load) corresponding to non-zero entries of $c$ are denoted as \emph{center buses}. Given an attacker vector $c$, the subgraph $\mathcal{S}$ controlled by the adversary is constructed using the following algorithm introduced in \cite{Hug2012}:
	\begin{enumerate}
		\item Let $\mathcal{S}$ be the set of all center buses.
		\item Extend $\mathcal{S}$ by including all branches and buses adjacent to center buses.
		\item If any bus on the boundary of $\mathcal{S}$ is a \textit{non-load bus} (\emph{i.e.}, no load is present), extend $\mathcal{S}$ by including all adjacent branches and buses to this bus.
		\item Repeat step 3) until all boundary buses are load buses.
	\end{enumerate}
	Constructing $\mathcal{S}$ with this method ensures that only measurements inside $\mathcal{S}$ can be modified by the attacker. The system operator will see the results of this unobservable attack as load changes at load buses within $\mathcal{S}$, while the total load of the system remain unchanged.

	\section{\label{sec:OldAttack}Worst-case Line Overflow Attacks}
	In \cite{Liang2015}, a bi-level optimization problem is introduced to find the worst-case unobservable line overflow attack vector given the attacker's limited resources to change states. In the optimization problem, the first level models the attacker's objective to maximize the power flow on a target line, subject to constraints on (1) the attacker's resources, characterized by the number of center buses in $c$, and (2) the attacker's detectability, characterized by the load shift, or the difference between the cyber load and the original load, as a percentage of the original load. The second level is a DCOPF problem simulating the system response to the attack. It is demonstrated in \cite{Liang2015} that unobservable attacks solved with the MILP can successfully lead to generation re-dispatch that maximize physical power flow on the target line, and hence, can result in line overflow for IEEE RTS 24-bus system. 
	
	In \cite{Liang2015}, $B$-$\theta$ method is used to fomulate the DCOPF problem, where the line power flow is calculated as the product of the dependency matrix of power flow and voltage angle $B$ and the voltage angle vector $\theta$. In contrast, in this paper we equivalently formulate the DCOPF using PTDF, where the line power flow is calculated as the product of PTDF matrix and power injection. Note that in this formulation, the variable vector $\theta$ is eliminated, and hence, the thermal limit constraints become independent of each other. Without loss of generality, we assume the power flow on the target line is positive. If it is not, we just maximize the negative of it. 
	
	The bi-level optimization problem is formulated as follows: (dual variables for second level problem are written in parentheses)
	\begin{flalign}
		\underset{c}{\text{maximize}}\:\; \hspace{0.17cm}& P_{l}-\sigma\left\Vert c\right\Vert _{1}\label{eq:Obj1_MaxPF}\\
		\notag \text{subject to}\hspace{0.2cm}\;\\
		& \hspace{-0.9cm}P=\text{PTDF}(G_{B}P_{G}^{*}-P_{D}) \label{eq:Physical_PF}\\
		& \hspace{-0.9cm}\left\Vert c\right\Vert _{1}\leqslant N_{1}\label{eq:con_resources}\\
		& \hspace{-0.9cm}-L_{S} P_{D}\leqslant Hc\leqslant L_{S} P_{D}\label{eq:con_loadshift}\\
		& \hspace{-0.9cm}\left\{P_{G}^{*}\right\} =\text{arg}\left\{ \underset{P_{G}}{\text{min}}\: C_{G}\left(P_{G}\right)\right\} \label{eq:OBJ_MINCOST}\\
		&\notag \hspace{-0.8cm} \text{subject to}\\
		&\hspace{-0.2cm}\begin{array}{lr} 
			\sum_{g=1}^{n_{g}}P_{Gg}=\sum_{i=1}^{n_{b}}P_{Di} & \hspace{0.8cm}(\lambda)\end{array}\label{eq:con_nodebalance}\\
		&\begin{array}{lc}
			\hspace{-0.2cm}-P^\text{max}\leqslant \text{PTDF}(G_{B}P_{G}-P_{D}+Hc)\\\hspace{2.5cm}\leqslant P^\text{max}  \hspace{0.9cm}(F^{\pm})\end{array}\label{eq:con_powerflow}\\
		& \begin{array}{cc}
			\hspace{-0.2cm} P_{G}^{\text{min}}\leqslant P_{G}\leqslant P_{G}^{\text{max}} &\hspace{1.6cm} (\alpha^{\pm})\end{array}\label{eq:con_GENlimit}
	\end{flalign}
	where the variables:
	\begin{description}[leftmargin=1.8cm,style=multiline]
		\item[$c$] is the $n_{b}\times1$ attack vector;
		\item[$P$] is the $n_{br}\times1$ vector of physical line power flow;
		\item[$P_l$] is the physical power flow of target line $l$;
		\item[$P_{G},P_{G}^{*}$] are $n_{g}\times 1$ vectors of generation dispatch variables and optimal
		generation dispatch solved by DCOPF, respectively;
		\item[$\lambda$] is the dual variable of the generation-load balance constraint;
		\item[$F^{+},F^{-}$] are $n_{br}\times1$ dual variable vectors of the upper and lower
		bound of thermal limits, respectively;
		\item[$\alpha^{+},\alpha^{-}$] are $n_{g}\times1$ dual variable vectors of the upper and lower bound
		of generator capacity limits, respectively;
	\end{description}
	and the parameters:
	\begin{description}[leftmargin=1.8cm,style=multiline]
		\item[$L_{S}$] is the load shift factor;
		\item[$P_{D}$] is the $n_{b}\times1$ vector of active load at each bus;
		\item[$N_{0}$] is the $l_{0}$-norm constraint integer;
		\item[$H$] is the $n_{b}\times n_{b}$ dependency matrix between power injection
		measurements and state variables;
		\item[$G_{B}$] is the $n_{b}\times n_{g}$ generator to bus connectivity matrix;
		\item[$C_{G}$] is the cost function of the generation vector;
		\item[$P^{\max}$] is the $n_{br}\times1$ vector of line thermal limit;
		\item[$P_{G}^{\min},P_{G}^{\max}$] are $n_{g}\times1$ vectors of minimum and maximum generator output,
		respectively;
		\item[$\sigma$] is the weight of the norm of attack vector $c$.
	\end{description}
	
	In the first level objective function, $\sigma$ is chosen to be a small positive number to minimize the contribution of the second term in the objective; constraint \eqref{eq:Physical_PF} calculates the physical power flow of the system; constraint \eqref{eq:con_resources} models the limited resources that the attacker can use to change states. Ideally, the  $l_0$-norm would be used, giving the exact sparsity of $c$, but for tractability we use the $l_1$-norm as a proxy. Constraint \eqref{eq:con_loadshift} ensures the load changes are small enough to avoid detection. In the second level, the system response to the attack vector determined in the first level is modeled via DCOPF as in \eqref{eq:OBJ_MINCOST}$-$\eqref{eq:con_GENlimit}. 
	
	The bi-level optimization problem introduced above is non-linear. We modify several constraints to convert the original formulation into an equivalent MILP as in \cite{Liang2015}. The modifications include:
	\begin{enumerate}
		\item Linearize the $l_1$-norm constraint in \eqref{eq:con_resources} by introducing a slack vector $s$ as
		\begin{flalign}
			c \leqslant s, \hspace{0.4cm} -c \leqslant s, \hspace{0.4cm} \underset{i\in \mathcal{L}_{\text{load}}}{\sum} s_{i} \leqslant N_{1} \hspace{0.4cm} \label{eq:l1-norm}
		\end{flalign}
		where $\mathcal{L}_{\text{load}}$ is the set of load buses. Meanwhile, the objective function \eqref{eq:Obj1_MaxPF} becomes
		\begin{flalign}
			\underset{c,s}{\text{maximize}}\:\; & P_{l}-\sigma \underset{i\in \mathcal{L}_{\text{load}}}{\sum}s_{i}\label{eq:Obj2_MaxPF}
		\end{flalign}
		\item Replace the second level DCOPF problem by its KKT optimality conditions as introduced in \cite{BoydBook}, as
		\begin{flalign}
			\notag\\\notag\\& \notag \hspace{-0.7cm}\eqref{eq:con_nodebalance} - \eqref{eq:con_GENlimit}\\
			\hspace{-0.2cm}\mathbf{0}=\hspace{0.1cm}& \notag \nabla \left[C_{G}(P_{G})\right]+\nabla (\sum_{g=1}^{n_{g}}P_{Gg}-\sum_{i=1}^{n_{b}}P_{Di})\cdot \lambda\\
			&\notag +\nabla \left[\text{PTDF} (G_{B}P_{G}-P_{D}+Hc)\mp P^{\text{max}}\right]\cdot F^{\pm}\\
			& +\nabla \left(P_{G}-P_{G}^\text{max}\right)\cdot \alpha^{+}+\nabla \left(P_{G}^\text{min}-P_{G}\right)\cdot \alpha^{-}\label{eq: DeriKKT}\\
			\mathbf{0} \leqslant & \hspace{0.1cm}F^{\pm}\label{eq:FGreater0}\\
			\mathbf{0} \leqslant & \hspace{0.1cm}\alpha^{\pm} \label{eq:AlphaGreater0}\\
			\mathbf{0}=& \hspace{0.1cm}\text{diag}(F^{\pm})\left[\text{PTDF} (G_{B}P_{G}-P_{D}+Hc)\mp P^{\text{max}}\right]\label{eq: CSFq}\\
			\mathbf{0}=&\hspace{0.1cm}\text{diag}(\alpha^{+})\left(P_{G}-P_{G}^\text{max}\right)\label{eq: CSalphap}\\
			\mathbf{0}=&\hspace{0.1cm}\text{diag}(\alpha^{-T})\left(P_{G}^\text{min}-P_{G}\right)\label{eq: CSalphan}
		\end{flalign}
		where constraint \eqref{eq: DeriKKT} is the partial gradient optimal condition, \eqref{eq:FGreater0} and \eqref{eq:AlphaGreater0} are the dual feasibility constraints, \eqref{eq: CSFq}--\eqref{eq: CSalphan} represent the complementary slackness constraints.
		\item Linearize the complementary slackness constraints in KKT conditions by introducing new binary variables $\delta$ and a large constant $M$ as
		\begin{flalign}
			& [\delta_{F}^{\pm};\delta_{\alpha}^{\pm}]\in \{0,1\}\label{eq:binary}\\
			\hspace{-0.6cm}& \hspace{-0.5cm} \left\{ \begin{array}{l} F^{\pm}\leqslant M\delta_{F}^{\pm} \\ P^{\text{max}}\mp \text{PTDF}(G_{B}P_{G}-P_{D}+Hc)\leqslant M(\mathbf{1}- \delta_{F}^{\pm})  \end{array}\right. \label{eq:Fbinary}\\
			\hspace{-0.6cm} & \hspace{-0.5cm} \left\{\begin{array}{l}\alpha^{\pm}\leqslant M\delta_{\alpha}^{\pm}\\
				P_{G}^{\text{max}}-P_{G}\leqslant M(\mathbf{1}-\delta_{\alpha}^{+})\\
				P_{G}-P_{G}^{\text{min}}\leqslant M(\delta_{\alpha}^{-}-\mathbf{1}) \end{array}\right.\label{eq:alphabinary}
		\end{flalign}
	\end{enumerate}
	
	The whole problem then becomes a single level MILP with objective \eqref{eq:Obj2_MaxPF}, subject to \eqref{eq:Physical_PF}, \eqref{eq:con_loadshift}, \eqref{eq:con_nodebalance}--\eqref{eq:l1-norm}, \eqref{eq: DeriKKT}--\eqref{eq:AlphaGreater0}, and \eqref{eq:binary}--\eqref{eq:alphabinary}. Throughout this paper, this problem is denoted as \emph{the original MILP} with $P_l^*$ as the optimal objective. This problem is NP-hard, thus it is not guaranteed that a solution can be found in polynomial time. As the system network size scales, the number of binary variables increases, resulting in an increased computational burden. We have found experimentally that for the IEEE 118-bus system, the original MILP fails to converge in a reasonable length of time using solver GUROBI. Since power systems in the real world are typically very large  (\emph{e.g.,} the PJM system includes 15000 buses, 2800 generators and 20000 branches), a straightforward approach to solving this problem for real-world systems does not allow a characterization of the worst-case attacks. 
	
	
	\section{\label{sec:Methods}Computational Efficient Algorithms to Solve Attack Optimization Problems}
	In this section, we introduce three computational efficient algorithms to overcome the computational challenges brought on by a large number of binary variables. Algorithm 1 (A1) reduces the number of line thermal limit constraints as well as the number of binary variables associated with them from the original MILP. If it converges, the optimal objective $P_{l}^{*(A1)}$ solved with Algorithm 1 is guaranteed to be equal to the optimal objective of the original MILP. However, as the system size scales, even though the number of binary variables associated with line thermal limit constraints is significantly reduced, the number of binary variables associated with generation limit constraints is still large enough to make the problem hard to solve. As an experimental verification, we note that Algorithm 1 works efficiently for the IEEE 118-bus system, but it fails to converge in a reasonable length of time for the Polish system with 2383 buses. Thus, similar to Algorithm 1, we propose Algorithm 2 (A2), in which we further reduce the number of generation limit constraints as well as the binary variables associated with them from the original MILP. Algorithm 2 gives a feasible solution, but the solution can be sub-optimal. Thus, the resulting objective value $P_{l}^{*(A2)}$ is a lower bound on $P_l^*$. Finally, we propose Algorithm 3, which maximizes the difference between the target line cyber and physical power flows, to evaluate system vulnerability without modeling the system response to the attack. This reduces to a linear program, whose optimal solution can be used to derive both a lower bound $P_l^{*(A3,\text{lb})}$ and an upper bound $P_l^{*(A3,\text{ub})}$ on $P_l^*$. A comparison of these three algorithms is given in Table \ref{tab:MethodComparison}.
	
	\subsection{\label{method1}Reducing the Number of Binary Variables Associated with Line Thermal Limit Constraints}
	
	In \cite{CAISO_SCUC} and \cite{ERCOT_SCED}, the authors suggest that a constraint reduction technique can be used to accelerate the solving process for optimization problems for large power systems, such as unit commitment and security constrained economic dispatch (SCED). The thermal limit constraints which are likely to be non-binding at the optimal solution are removed for speed. In Algorithm 1, we apply a similar method to improve the efficiency of the original MILP by reducing the number of binary variables associated with line thermal limits constraints. For lines whose pre-attack power flow are much lower than their ratings, their thermal limit constraints are unlikely to be binding in the optimal solution of the optimization problem, and hence, can be removed. Thus, we propose Algorithm \ref{alg:Method1} to remove constraints from \eqref{eq:binary} and \eqref{eq:Fbinary} without violating any constraints of the original MILP.
	\begin{algorithm}[tbh]
		\protect\caption{Attack optimization problem with reduced thermal limit constraints}	\label{alg:Method1}			
		\begin{enumerate}
			\item Perform a system wide DCOPF assuming no attack is present.
			\item Let $\mathcal{Q}$ be the set of lines whose power flow are more than 90\% of ratings (denote as \emph{critical lines}). 
			\item For each line $k$, if $k \notin \mathcal{Q}$, remove the corresponding constraints from \eqref{eq:con_powerflow}, \eqref{eq: DeriKKT}--\eqref{eq:FGreater0}, \eqref{eq:binary}, and \eqref{eq:Fbinary}.
			\item Solve the reduced problem, and use the optimal dispatch $P_G^{*(A1)}$ to calculate the cyber power flow of the system. 
			\item If there exists any cyber overflow, add those lines to $\mathcal{Q}$, and go back to 3).
			\item Let $P_l^{*(A1)}$ be the optimal objective value of the reduced problem. 
		\end{enumerate} 
	\end{algorithm}
	
	
	If the algorithm terminates, the solution is guaranteed to be the optimal solution of the original MILP (\emph{i.e.} $P_l^{*(A1)}=P_l^*$) because no thermal limits are violated. 
	
	
		\begin{table*}[t]
			\renewcommand{\arraystretch}{1.3}
			\protect\caption{Comparison
				of Three Proposed Algorithms\label{tab:MethodComparison}}
			\centering
			\vspace{-0.2cm}
			\begin{tabular}{c|c|c|>{\centering}m{4cm}|c|c|c|c}
				\thickhline
				\multirow{2}{*}{Algorithm} & \multirow{2}{*}{Type of Program} & \multirow{2}{*}{Outcome} & \multirow{2}{*}{Tractable Test Cases}& \multicolumn{2}{c|}{\# of Binary Variables} & \multicolumn{2}{c}{Computation Time (s)}\tabularnewline
				\hhline{~~~~----}
				& & &  & 118-bus & Polish & 118-bus & Polish \tabularnewline
				\hline
				Original MILP & MILP & Optimal Solution & 24-bus & 480 & 6446 & - & - \tabularnewline
				\hline 
				Algorithm 1 & MILP & Optimal Solution & 24-bus, 118-bus& 122 & 688 & 22.04 & - \tabularnewline
				\hline 
				Algorithm 2 & MILP & Lower Bound & 24-bus, 118-bus, Polish (2383-bus)& 45 & 87 & 1.03 & 40.41 \tabularnewline
				\hline 
				Algorithm 3 & LP & Lower \& Upper Bounds & 24-bus, 118-bus, Polish (2383-bus) & - & - & 0.35 & 4.57  \tabularnewline
				\thickhline 
			\end{tabular}
		\end{table*}

	\subsection{\label{method2}Reducing the Number of Binary Variables Associated with Power Generation Limit Constraints} 
	Algorithm 2 modifies Algorithm 1 by further reducing the number of binary variables, now focusing on generation limit constraints. Since constraint \eqref{eq:con_loadshift} ensures the load changes are small, the generation re-dispatch is likely to be limited within a small number of generators (denote as \emph{marginal generators}) corresponding to these load changes. Algorithm 2 reduces the number of binary variables associated with generators by assuming the generation level of non-marginal generators remains unchanged after attack.
	
	\begin{algorithm}[tbh]
		\protect\caption{Attack optimization problem with reduced thermal limit \& generation limit constraints}	\label{alg:Method2}			
		\begin{enumerate}
			\item Perform a system wide DCOPF assuming no attack is present.
			\item Let $\mathcal{Q}$ be the set of critical lines as defined above.
			\item Let $\mathcal{R}$ be the set of marginal generators. \emph{i.e.}, the set of generators $g$ where $P_{Gg}^{\text{min}} < P_{Gg} < P_{Gg}^{\text{max}}$.
			\item For each line $k$ and each generator $g$, if $k \notin \mathcal{Q}$ and $g \notin \mathcal{R}$, remove the corresponding constraints from \eqref{eq:con_powerflow}--\eqref{eq:con_GENlimit}, \eqref{eq: DeriKKT}--\eqref{eq:AlphaGreater0} and \eqref{eq:binary}--\eqref{eq:alphabinary}.
			\item Solve the reduced problem, then take the optimal attack vector $c_{A2}^*$ to run post-attack DCOPF \eqref{eq:OBJ_MINCOST}--\eqref{eq:con_GENlimit} to get system operator's corresponding dispatch $P_G^{\text{post}}$. 
			\item If $P_G^{*(A2)} \neq P_G^{\text{post}}$, add those generators with different dispatch to the set $\mathcal{R}$, and then go back to 4).
			\item Take $P_G^{*(A2)}$ to calculate the cyber power flow of the system.
			\item If there exists any cyber overflow, add those lines to $\mathcal{Q}$, and go back to 4).
			\item Let $P_l^{*(A2)}$ be the optimal objective value of the reduced problem.
		\end{enumerate}
	\end{algorithm}
	
	By significantly reducing the number of binary variables, Algorithm 2 can be efficiently applied to the Polish system. Since some of the variables $P_G$ are held constant, the solution is not guaranteed to be optimal for the original MILP. However, it does provide a feasible solution, and hence, $P_{l}^{*(A2)}$ is a lower bound on $P_l^*$.

	\subsection{\label{method3}Maximizing the Difference between Post-attack Physical and Cyber Power Flows}
	
	In this section, we propose Algorithm 3 that provides both lower and upper bounds on $P_l^*$. In the original MILP the objective is to maximize the physical power flow of a target line, which is calculated by \eqref{eq:Physical_PF}. Substituting \eqref{eq:Physical_PF} into \eqref{eq:con_powerflow} gives 
	\begin{flalign}
		-P^\text{max}-\text{PTDF}(Hc)\leqslant P\leqslant P^\text{max}-\text{PTDF}(Hc).\label{eq:upperbound}
	\end{flalign}
    In particular, for target line $l$, 
	\begin{flalign}
		P_l\leqslant P_l^\text{max}-\text{PTDF}^l(Hc) \label{whyA3ub}
	\end{flalign}
	where $P_l^\text{max}$ is a constant and $\text{PTDF}^l$ is the $l^\text{th}$ row of the PTDF matrix. Thus, we can maximize $-\text{PTDF}^l(Hc)$ to find an upper bound on $P_l^*$ constrained by \eqref{eq:Physical_PF}, \eqref{eq:con_loadshift}, and \eqref{eq:l1-norm}. This optimization problem is formulated as  
	\begin{flalign}
		\underset{c,s}{\text{maximize}}\:\; & -\text{PTDF}^l(Hc)\label{eq:Obj3_MaxDiff}\\
		\notag \text{subject to} \hspace{0.2cm}& \eqref{eq:Physical_PF}, \eqref{eq:con_loadshift},  \eqref{eq:l1-norm}	
	\end{flalign}
	This optimization is a linear program and can be solved efficiently for large systems.
	\begin{algorithm}[tbh]
		\protect\caption{Maximizing the difference between post-attack physical and cyber power flows}	\label{alg:Method3}			
		\begin{enumerate}
			\item Solve optimization problem \eqref{eq:Obj3_MaxDiff}. Let $c_{A3}^*$ be the resulting optimal attack vector.
			\item Calculate the upper bound as 
			\begin{flalign}
				P_l^{*(A3,\text{ub})} = P_l^\text{max}-\text{PTDF}^l(Hc_{A3}^*).
			\end{flalign}
			\item Perform post-attack DCOPF \eqref{eq:OBJ_MINCOST}--\eqref{eq:con_GENlimit} with $c=c_{A3}^*$ to find dispatch $P_{G}^{\text{post}}$.
			\item Calculate the lower bound as
			\begin{flalign}
				P_l^{*(A3,\text{lb})} = \text{PTDF}^l(G_{B}P_{G}^{\text{post}}-P_{D}).\label{lba3}
			\end{flalign}
		\end{enumerate}
	\end{algorithm}
	
	Once the optimal objective is reached, $P_l^{*(A3,\text{ub})}$, can be calculated by adding this optimal objective to $P_l^\text{max}$ according to \eqref{whyA3ub}.
	The system re-dispatch $P_G^{\text{post}}$ and line cyber power flow $P^{\text{cyber}}$ are achieved by performing post-attack DCOPF \eqref{eq:OBJ_MINCOST}--\eqref{eq:con_GENlimit}. The resulting physical power flow on the target line, $P_l^{*(A3,\text{lb})}$, is a lower bound on $P_l^*$.
	
	Note that if the load shift is too large, it is possible that the post-attack DCOPF fails to converge. However, since we keep the load shift small, this situation is beyond the scope of this paper. Furthermore, if $P_l^{*(A3,\text{ub})}=P_l^{*(A3,\text{lb})}$, they are the optimal objective of the original MILP. This is the situation when $\left|P_l^{\text{cyber}}\right|=P_l^\text{max}$, \emph{i.e.}, \eqref{eq:con_powerflow} holds with equality in the optimal solution of post-attack DCOPF \eqref{eq:OBJ_MINCOST}--\eqref{eq:con_GENlimit}.

	
	\section{\label{sec:Simulation}Simulation results}
	In this section, we apply algorithms described in Sec. \ref{sec:Methods} to both the IEEE 118-bus system and the Polish system. As stated in Sec. \ref{sec:Methods}, Algorithm 1 does not converge in reasonable length of time for the Polish system. Therefore, only Algorithms 2 and 3 are applied on the Polish system. There are 7 critical lines prior to the attack in IEEE 118-bus system and 17 in the Polish system. The number of marginal generators prior to the attack for the two systems are 15 and 6, respectively. We exhaustively evaluate vulnerability of these two systems by targeting all critical lines with load shift constraint $L_S=$ 10\%. The $l_1$-norm constraint $N_1$ is chosen as $\left[0.1,1\right]$ for the 118-bus system and $\left[0.1,2\right]$ for the Polish system, both with increment 0.1. GUROBI is utilized as the solver to solve the optimization problem. Throughout, we use MATLAB R2014a and MATPOWER package v4.1 to perform the simulation. 
	

	
	\subsection{\label{Efficiency}Computational Efficiency} 	
	The computational efficiency improved by Algorithms 1 and 2 are characterized as the reduction in the number of binary variables. In Table \ref{tab:MethodComparison}, the $5^{\text{th}}$ column compares the average numbers of binary variables of the original MILP, Algorithm 1, and Algorithm 2 for both test systems. Note that for the Polish system, the number of binary variables of Algorithms 1 is theoretical, as Algorithm 1 does not converge in a reasonable length of time for the Polish system. The $6^{\text{th}}$ column summarizes the average computation time of the three methods for two typical test cases, \textit{i.e.,} test cases with target line 104 in 118-bus system and target line 292 in the Polish system. As one would expect, Algorithm 3, an LP, is computationally the most efficient among the three proposed algorithms for both test systems. 

	\subsection{\label{118-bus}Results for the IEEE 118-bus System} 
\begin{figure}[h]
		\centering{}\includegraphics[trim=0 0.3cm 0 0.3cm, scale=0.76]{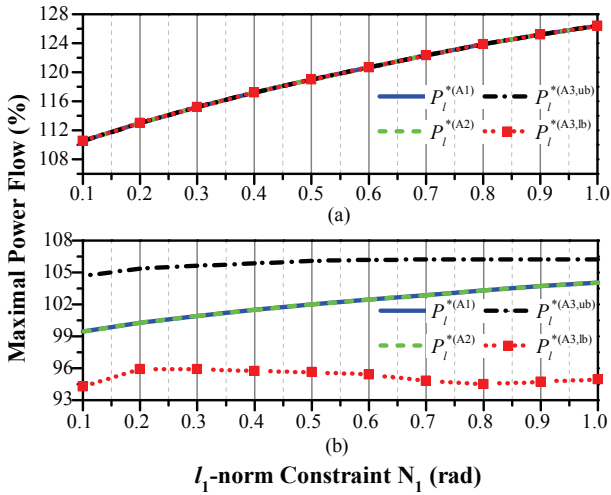}\protect\protect\caption{The maximum power flow v.s. the $l_1$-norm constraint ($N_{1}$) when target lines are (a) line 104, and (b) line 141 of IEEE 118-bus system. \label{fig:118_t104}}
	\end{figure}

	Figs. \ref{fig:118_t104}(a)	and (b) illustrate the maximum power flow on target lines 104 and 141, respectively, in IEEE 118-bus system using the three different algorithms described in Sec. \ref{sec:Methods}. We use this system to compare the bounds of Algorithms 2 and 3 to the exact solution provided by Algorithm 1; note, however, that the optimal solution of Algorithm 1 is not available for the Polish system. Both figures illustrate that as $N_1$ increase, $P_l^*$ also increases, and all three algorithms can result in overflow.
	
	Fig. \ref{fig:118_t104}(a) shows that for target line with any $N_1$, $P_l^{*(A1)}=P_l^{*(A2)}=P_l^{*(A3,\text{ub})}=P_l^{*(A3,\text{lb})}=P_l^{*}$; that is, all three algorithms achieve the optimal solution of the original MILP. Fig. \ref{fig:118_t104}(b) shows that for target line 141,  $P_l^{*(A3,\text{lb})}<P_l^{*(A1)}=P_l^{*(A2)}<P_l^{*(A3,\text{ub})}$, illustrating that $P_l^{*(A3,\text{lb})}$ and $P_l^{*(A3,\text{ub})}$ are not always tight bounds on $P_l^{*}$. In all scenarios we have considered, Algorithm 2 yields the optimal solution for the 118-bus system. (This is not true for the Polish system, as illustrated in Fig. \ref{fig:Polish_t4}.) 
	\subsection{\label{Polish}Results for the Polish System}
\begin{figure}[h]
		\centering{}\includegraphics[trim=0 0.3cm 0 0.3cm, scale=0.76]{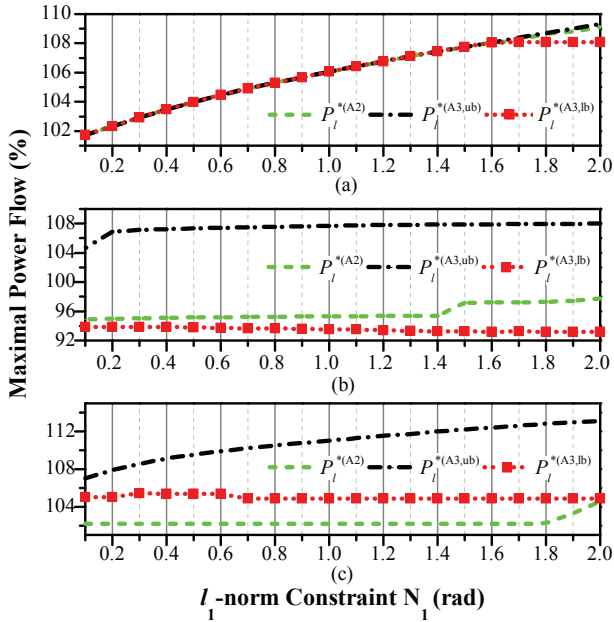}\protect\protect\caption{The maximum power flow v.s. the $l_1$-norm constraint ($N_{1}$) when target lines are (a) line 292, (b) line 4, and (c) line 1816 of the Polish system.. \label{fig:Polish_t4}}
	\end{figure}

	Results for target lines 292, 4, and 1816 obtained with Algorithms 2 and 3 are illustrated in Figs. \ref{fig:Polish_t4}(a), (b), and (c), respectively. We observe that for target line 292, the upper and lower bounds exactly match, \textit{i.e.,} $P_l^{*(A3,\text{ub})}=P_l^{*(A3,\text{lb})}=P_l^{*(A2)}$, in the range $N_1\in\left[0.1,1.6\right]$, and therefore, the optimal solutions are reached. For the remaining cases, our algorithms do not give the optimal solutions, since $P_l^{*(A3,\text{lb})}<P_l^{*(A2)}<P_l^{*(A3,\text{ub})}$. For target line 4, we observe that the upper and lower bounds do not match, but the lower bound from Algorithm 2 is tighter than that from Algorithm 3, whereas for target line 1816, the opposite is true. 
	
	

	\section{\label{sec:conclusion}Conclusion}
	This paper presented an approach to evaluate the vulnerability of power systems to FDI attacks irrespective of the size of the network. Three computationally efficient algorithms have been introduced to characterize the vulnerability of systems. It is demonstrated that Algorithm 1 can find worst-case attacks. Algorithms 2 and 3, on the other hand, can find feasible attacks that can result in significant consequences. Future work will include evaluating attack efficiency for arbitrary system size when attacker only has limited information as in \cite{JZhangGM16} using the three algorithm proposed here. In addition, designing detection countermeasures to thwart such attacks is also of interest.

	\section*{Acknowledgment}
	This work is supported jointly by the National Science Foundation and the Department of Homeland Security under Grant CNS-1449080.
	

	%
	%

	
	
	%
	%
	%
	
	\bibliographystyle{IEEEtran}
	\bibliography{DistributedAttacks}

\end{document}